\pdfoutput=1
\documentclass[aps,pre,twocolumn,amsmath,amssymb,showpacs,mathptmax]{revtex4-1}
\usepackage{graphicx,color,hyperref}

\usepackage{mathptmx}
\usepackage{times}
\usepackage{epsfig,amsopn}
\usepackage{graphicx}
\usepackage{bm,amsmath,amssymb}
\usepackage{natbib}
\usepackage{braket}
\usepackage{float}
\usepackage{enumerate}
\usepackage{hyperref}
\allowdisplaybreaks[1]

\begin{document}
\def\be{\begin{equation}}
\def\ee{\end{equation}}
\def\bea{\begin{eqnarray}}
\def\eea{\end{eqnarray}}
\def\fr{\frac}
\def\l{\label}

\newcommand{\eref}[1]{Eq.~(\ref{#1})}%
\newcommand{\Eref}[1]{Equation~(\ref{#1})}%
\newcommand{\fref}[1]{Fig.~\ref{#1}} %
\newcommand{\Fref}[1]{Figure~\ref{#1}}%
\newcommand{\sref}[1]{Sec.~\ref{#1}}%
\newcommand{\Sref}[1]{Section~\ref{#1}}%
\newcommand{\aref}[1]{Appendix~\ref{#1}}%
\newcommand{\sgn}[1]{\mathrm{sgn}({#1})}%
\newcommand{\erfc}{\mathrm{erfc}}%

%%%%%%%%%%%%%%%%%%%%%%%%%%%%%%%%%%%%%%%%%%%%%%%%%%%%%%%
\title{Diffusion in a potential landscape with stochastic resetting}
\author{Arnab Pal}
\affiliation{Raman Research Institute, Bangalore 560080, India}
%\email{arnab@rri.res.in}

\date{\today}

\begin{abstract}
The steady state of a Brownian particle
diffusing in an arbitrary potential
under the stochastic resetting mechanism  
has been studied.
We show
that there are different classes of nonequilibrium steady states depending
on the nature of the potential.
In the stable potential landscape,
the system attains a well defined steady state
however existence of the steady state
for the unstable landscape is constrained.
We have also investigated the 
transient properties of the propagator towards the steady state
under the
stochastic resetting mechanism. Finally, we have done numerical simulations
to verify our
analytical results.

% It is well known that a free Brownian
%particle reaches a steady state when subjected to stochastic 
%resetting. We have extended the study for a particle 
%diffusing in a general potential landscape. 
%
%%%%%%%%%%%%%%%%%%%
\end{abstract}

\maketitle

%%%%%%%%%%%%%%%%%%%%%%%%%%%%%%%%%%%%%%%%%%%%%%%%%%%%%%%

\section{Introduction}
\label{sec:Introduction}
Diffusion
with stochastic resetting is considered
to be a natural framework
for the study of intermittent search processes
\cite{Bell-91, Benichou-11}.
The simplest question of finding a lost object being a key or a car
or an offender is one of the central quests of the discipline.
The search processes related to resetting
are realized in diverse fields such as biochemistry (where
a signaling molecule is reset back to a receptor protein
in the membrane depending on
the concentration of certain molecules in the vicinity) \cite{Adam-68},
computer network (to find an element in a sorted and pivoted array)
\cite{Montanari-02},
ecology (Capuchin monkeys, known for long
term memory, foraging a territory with palm nuts) \cite{Bartumeus-09}
and microbiology \cite{Cates-12}.
In addition,
this mechanism was considered to compute the stationary distribution
of variant models of population growth where the population
is stochastically reset to some higher or lower values
leading to a power law growth \cite{Manrubia-99, Visco-10}.
Also, there have been interests to study the continuous time
random walk where the both the position and the waiting time
are chosen from certain distributions in the presence of 
resetting \cite{Montero-13}.

`Stochastic resetting' is a mechanism
where a Brownian particle is stochastically reset to its initial
position at a constant rate thus driving
the system away from any equilibrium state
\cite{Evans:2011-1, Evans:2011-2, Evans:2013, Whitehouse:2013}.
It is thus a
simple mechanism to generate a non equilibrium
stationary state. In such states probability currents are
non-zero and detailed balance
does not hold naturally.
Of late, the implication of the stochastic reset
has been studied in the one dimensional reaction-diffusion systems where
a finite reset rate leads to an unique non equilibrium stationary state \cite{Henkel:2014}.
The interface growth models described by Kardar-Parisi-Zhang and Edwards-Wilkinson
equations also exhibit
nonequilibrium stationary states with non-Gaussian interface fluctuations
when the interface stochastically resets to
a fixed initial profile at a constant rate \cite{Gupta:2014}.
In this backdrop, a natural question to ask
would be: is the non equilibrium stationary state
generic to any dynamics subjected to stochastic resetting.
The primary goal of this paper is to address this question.
To gain insights, one considers model systems which are
simple enough though addresses the basic moral.
In this paper, we consider
a simple model of a Brownian particle
diffusing in an arbitrary potential landscape in the
presence of stochastic resetting.
It is obvious that for a bounded case, even without reset one gets
a steady state around the minimum of the potential.
But when the equilibrium point of the potential
differs from the reset point, two mechanisms compete with each
other and finally reaches a steady state which shows certain generic
behavior.
On the other hand, for a particle diffusing in an unbounded potential,
there exists no steady state at all in the absence of resetting.
We propose to invoke stochastic resetting to retrieve the
steady state. However, this behavior is not universal and rather puts a
general constraint on the nature of the potential.
We derive the conditions that
ensures the steady state in the case of an unbounded potential.

The paper is structured as follows. In the following section, we
introduce the model and the resetting dynamics. In
\sref{steady-state}, we obtain exact steady state distribution $P_{\rm
st}(x|x_0)$ for two representative choices of the potential $V(x)$, namely,
(i) $V(x) \sim  \mu |x|$, and (ii) $V(x) \sim \mu x^2$.
The positive and negative sign of $\mu$ describe bounded and
unbounded potential respectively. We also derive the
conditions to obtain an unique steady state for
an arbitrary potential landscape. In \sref{transient},
we investigate the transient behavior of the propagator
in the presence of the resetting. We conclude with a
summary and future directions in \sref{summary}.
\\
%%%%%%%%%%%%%%%%%%%%%%%%%%%%%%%%%%%%%%%%%%%%%%%%%%%%%%%
\section{The model}
\l{reset-dynamics}
Consider a single particle undergoing diffusion in one dimension
in presence of an external potential $V(x)$:
\be
\frac{dx}{dt}=-V'(x)+\eta(t),
\l{eq:eom}
\ee
where $\eta(t)$ is a Gaussian white noise with
\be
\langle\eta(t)\rangle=0,
~~\langle\eta(t)\eta(t')\rangle=2D\delta(t-t'),
\ee
$D$ being the diffusion constant and the viscosity of the medium has been scaled to unity for brevity.
Here, angular brackets denote averaging over noise realizations.
The initial condition is 
\be
x(0)=x_{0} ,
\ee
where $x_0 \in (0,\infty]$.
We now introduce the `Stochastic resetting' mechanism by which the particle returns
to its initial location at a constant rate $r$. To elaborate, in a
small time $\Delta t,$ the particle is reset to the initial position $x=x_0$ with probability
$r\Delta t$, while with the complementary probability $1-r\Delta t,$
the particle dynamics follows \eref{eq:eom}.

%%%%%%%%%%%%%%%%%%%%%%%%%%%%%%%%%%%%%%%%%%%%%%%%%%%%%%%
\section{Steady state distribution}
\l{steady-state}
Let $P(x,t|x_0)$ be the probability to find the particle at position $x$
at time $t$, given that it was at $x_0$ at time $t=0$.
From the dynamical rules for the evolution of the particle given in the
preceding section, it follows 
that 
\bea
\frac{\partial P}{\partial
t}=D\frac{\partial^{2}P}{\partial
x^{2}}+\frac{\partial [V'(x)P]}{\partial
x}-rP+r\delta(x-x_0),
\l{transient}
\eea
with the initial condition $P(x,0|x_0)=\delta(x-x_{0})$. Here, the third and
fourth terms on the right hand side (rhs) account for the resetting
events, denoting the negative probability flux $-rP$ from each point $x$ and a
corresponding positive probability flux into $x=x_0$. The steady
state solution $P_{\rm st}(x|x_0)$ satisfies
\be
0=D\frac{d^{2}P_{\rm st}}{d
x^{2}}+\frac{d [V'(x)P_{\rm st}]}{d x}-rP_{\rm st}+r\delta(x-x_0).
\l{eq:steady-state}
\ee
In the following section, we have investigated steady state distributions
for various bounded and unbounded potential landscapes.
In particular, we have studied for two representative choices of the potential $V(x)$, namely,
(i) $V(x) \sim  \mu |x|$, and (ii) $V(x) \sim \mu x^2$.

%%%%%%%%%%%%%%%%%%%%%%%%%%%%%%%%%%%%%%%%%%%%%%%%%%%%%%%
%%%%%%%%%%%%%%%%%%%%%%%%%%%%%%%%%%%%%%%%%%%%%%%%%%%%%%%
\subsection{The case of a mod potential}
We first consider the case of a mod potential. This potential
is centred either around its minimum or the maximum at $0$.
The reset location is at $x_0 \ne 0$. The nature of the
potential allows us
to identify three regions in $x$, namely, region I
$(x>x_0)$, region II $(0<x<x_0)$, and region III $(x<0)$. To
find the steady state,
we solve
Eq.~(\ref{eq:steady-state}) in each region and require that the solutions
are continuous at $x=x_0$ and $x=0$. Though, the derivatives are discontinuous
and it can be seen by integrating
Eq.~(\ref{eq:steady-state}) over an infinitesimal region
around $x=x_0$,
\bea
\fr{dP_{\rm st}^I(x|x_0)}{dx}\Big|_{x=x_0}-\fr{dP_{\rm st}^{II}(x|x_0)}{dx}\Big|_{x=x_0}=-\fr{r}{D},
\label{MC1}
\eea
This discontinuity does not depend on $\mu$ indicating
the robustness of `kinks' present at $x_0$ irrespective of potential landscapes.
On the other hand, while integrating Eq.~(\ref{eq:steady-state}) over an infinitesimal region
around $x=0$, we find that
\bea
\fr{dP_{\rm st}^{II}(x|x_0)}{dx}\Big|_{x=0}-\fr{dP_{\rm
st}^{III}(x|x_0)}{dx}\Big|_{x=0}=\mp \fr{2\mu}{D}P_{\rm
st}^{II}(x|x_0)\Big|_{x=0},
\label{MC2}
\eea
in which minus and plus signs are for the bounded and the unbounded case respectively.
In the following subsections, we consider these two cases respectively.

%%%%%%%%%%%%%%%%%%%%%%%%%%%%%%%%%%%%%%

\subsubsection{Bounded potential:~$V(x)=\mu |x|,~\mu>0$}
We first consider the case where $\mu>0$.
Using the \eref{MC1} and \eref{MC2}, we obtain the steady state solutions
given by,
\bea
P^{I}_{\rm st}(x|x_{0})&=&\fr{r}{\sqrt{\mu^{2}+4Dr}}e^{-m_{2}x_{0}}e^{m_{2}x}\nonumber \\
&&+\fr{\mu r}{\sqrt{\mu^{2}+4Dr}(\sqrt{\mu^{2}+4Dr}-\mu)}e^{-m_{1}x_{0}}e^{m_{2}x},\nonumber \\ 
P^{II}_{\rm st}(x|x_{0})&=&\fr{r}{\sqrt{\mu^{2}+4Dr}}e^{-m_{1}x_{0}}e^{m_{1}x}\nonumber \\
&&+\fr{\mu r}{\sqrt{\mu^{2}+4Dr}(\sqrt{\mu^{2}+4Dr}-\mu)}e^{-m_{1}x_{0}}e^{m_{2}x},\nonumber \\
P^{III}_{\rm st}(x|x_{0})&=&\fr{ r}{\sqrt{\mu^{2}+4Dr}-\mu}e^{-m_{1}x_{0}}e^{-m_{2}x},
\l{PDF-mod-bound}
\eea
where
\bea
 m_{1}=\fr{-\mu+\sqrt{\mu^{2}+4Dr}}{2D},~
 m_{2}=-\fr{\mu+\sqrt{\mu^{2}+4Dr}}{2D}.
\label{m1,m2}
\eea
\\
\fref{fig1} shows a comparison between simulations and theory for
steady state \eref{PDF-mod-bound} demonstrating a very good agreement. From
the solution, it is evident that $P_{\rm
st}(x|x_0)$ exhibits two cusps where its derivatives are discontinuous,
namely, (i) at the resetting location $x=x_0$, and (ii) at $x=0$, the
point at which the potential $V(x)$ has discontinuous derivatives. 

%%%%%%%%%%%%%%%%%%%%%%%%%%%%%%%%%
\begin{figure*}
\includegraphics[width=.3\hsize]{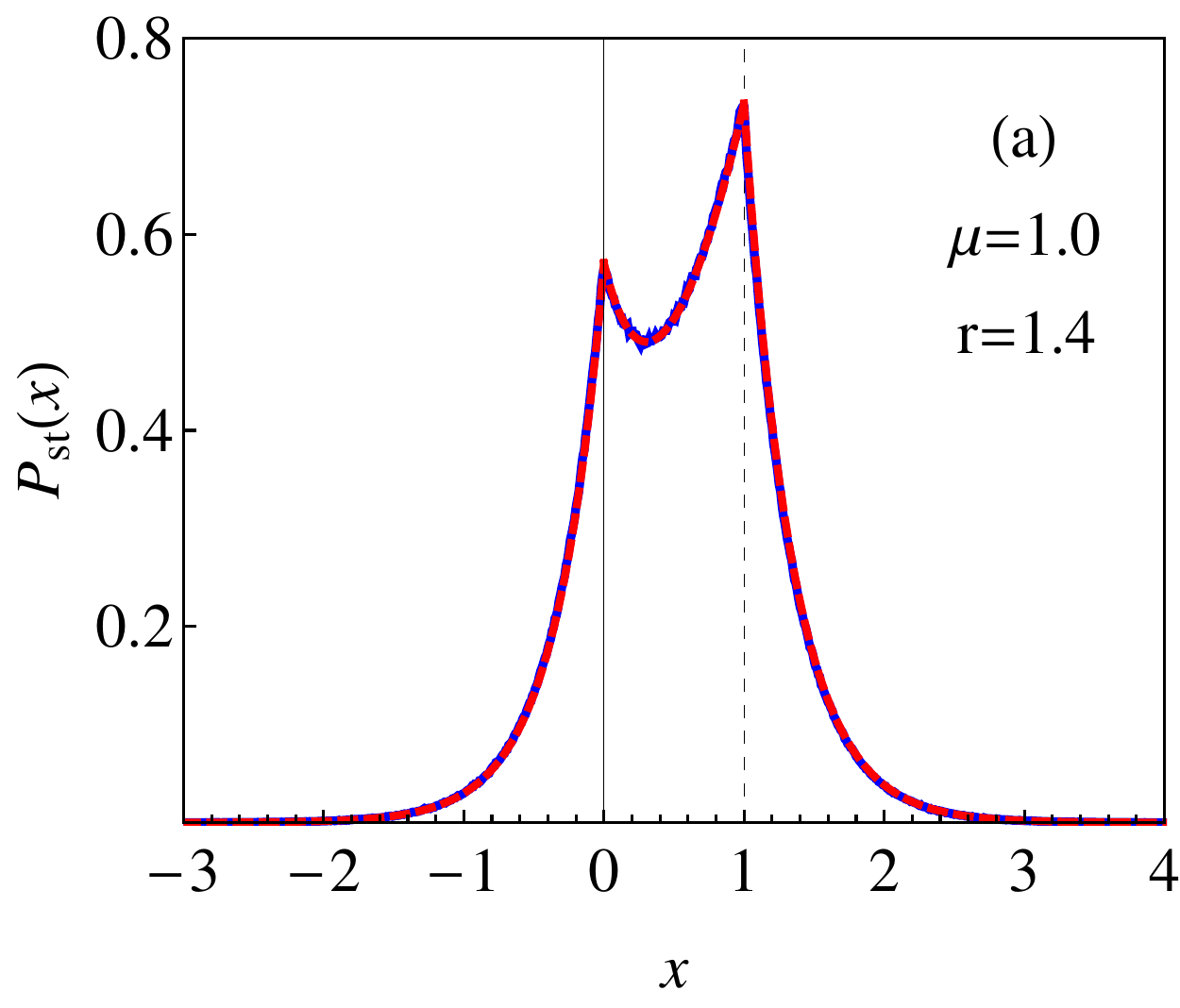}
\includegraphics[width=.3\hsize]{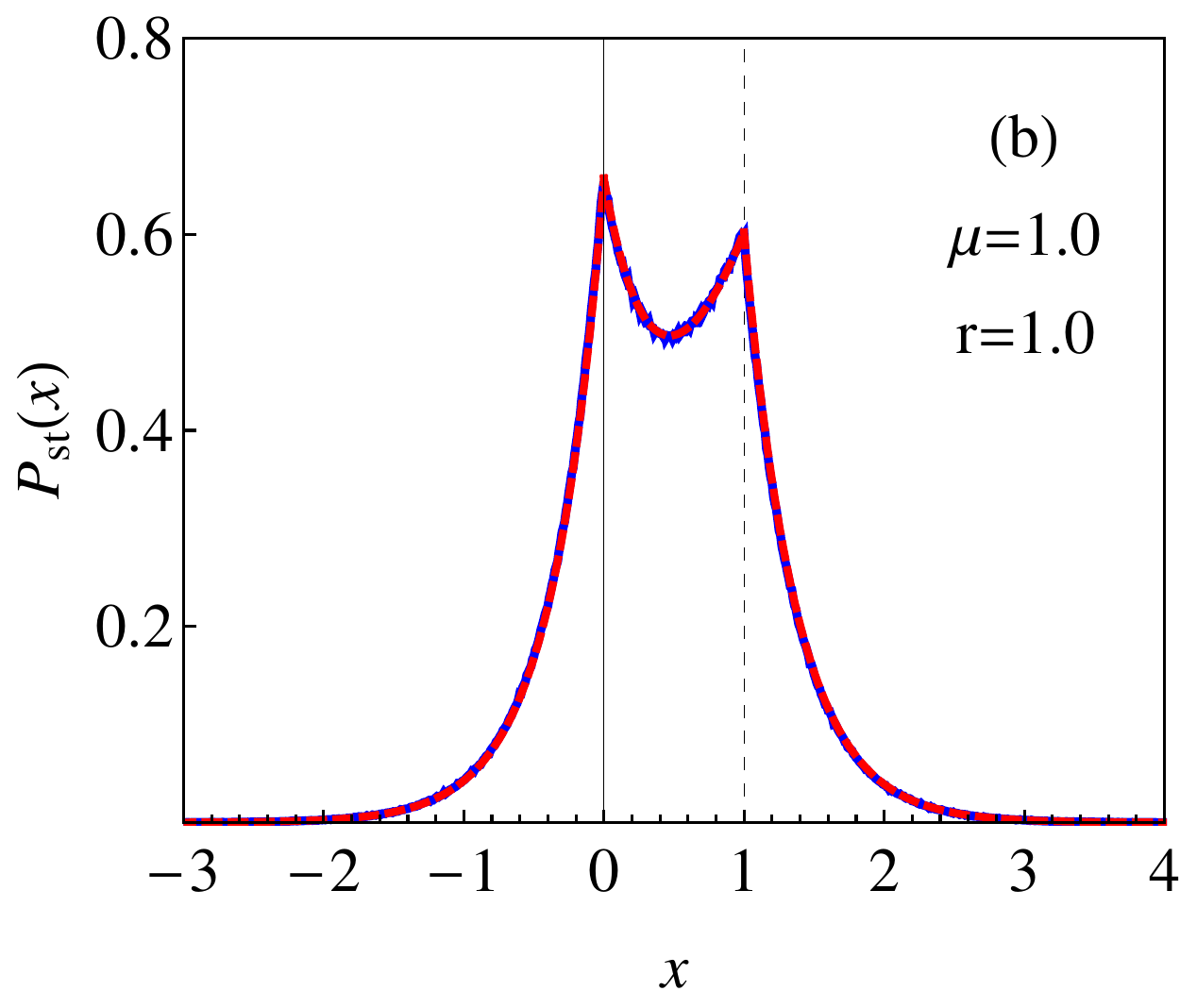}
\includegraphics[width=.3\hsize]{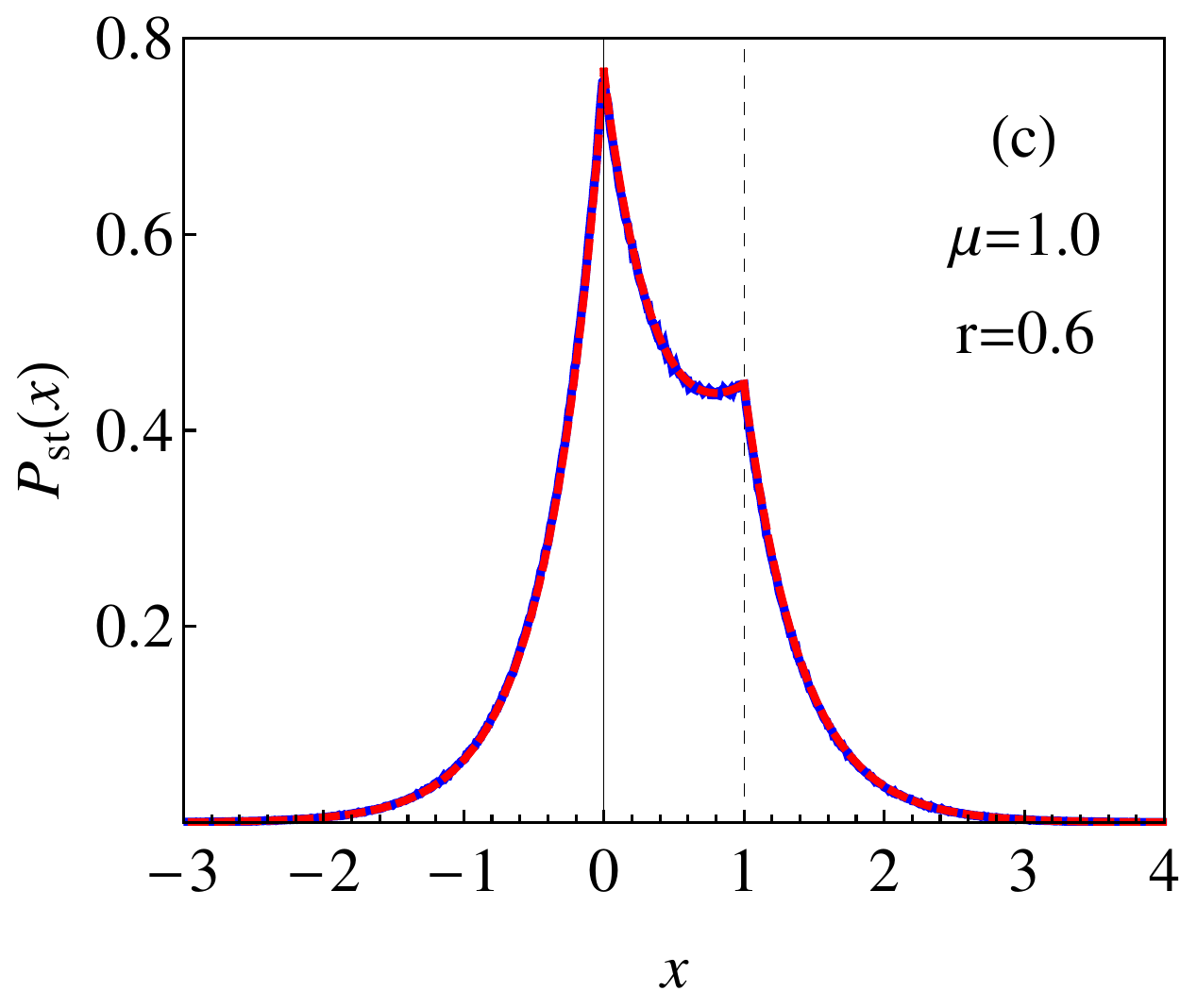}
\caption{(Color online) Stationary distribution $P_{\rm st}(x|x_0)$ for
the bounded potential $V(x)=\mu|x|$, with $\mu >0$.
We choose $D=0.5,x_0=1.0,\mu=1.0$ while vary $r$. The (red) dashed
line plots the analytical result for $P_{\rm st}(x)$, while the (blue)
points are numerical simulation results. Also,
The vertical solid and dashed lines indicate the location of the
stable minimum of the bounded potential and the reset point $x_0$ respectively.
The motion of the peak is also clear from the figure.}
\l{fig1}
\end{figure*}

%%%%%%%%%%%%%%%%%%%%%%%%%%%%%%%
\subsubsection{Unbounded potential:~$V(x)=\mu|x|,~\mu<0$}
The steady state solutions for the unbounded case when $\mu<0$ are given
in the following,
\bea
P^{I}_{\rm st}(x|x_{0})&=&\fr{r}{\sqrt{\mu^{2}+4Dr}}e^{m_{1}x_{0}}e^{-m_{1}x}\nonumber\\
&&-\fr{\mu r}{\sqrt{\mu^{2}+4Dr}(\sqrt{\mu^{2}+4Dr}+\mu)}e^{m_{2}x_{0}}e^{-m_{1}x},\nonumber \\ 
P^{II}_{\rm st}(x|x_{0})&=&\fr{r}{\sqrt{\mu^{2}+4Dr}}e^{m_{2}x_{0}}e^{-m_{2}x}\nonumber \\
&&-\fr{\mu r}{\sqrt{\mu^{2}+4Dr}(\sqrt{\mu^{2}+4Dr}+\mu)}e^{m_{2}x_{0}}e^{-m_{1}x},\nonumber \\
P^{III}_{\rm st}(x|x_{0})&=&\fr{ r}{\sqrt{\mu^{2}+4Dr}+\mu}e^{m_{2}x_{0}}e^{m_{1}x},
\l{PDF-mod-unbound}
\eea
where $m_{1},~m_{2}$ are given by \eref{m1,m2}.
\\
\fref{fig2} shows a comparison between simulations and theory for
steady state \eref{PDF-mod-unbound} demonstrating a very good agreement.
Again $P_{\rm st}(x|x_0)$ exhibits two cusps where
its derivatives are discontinuous,
namely, (i) at the resetting location $x=x_0$, and (ii) at $x=0$, the
point at which the potential $V(x)$ has discontinuous derivatives. 

%%%%%%%%%%%%%%%%%%%%%%%%%%%%%%%%%
\begin{figure*} 
\includegraphics[width=.3\hsize]{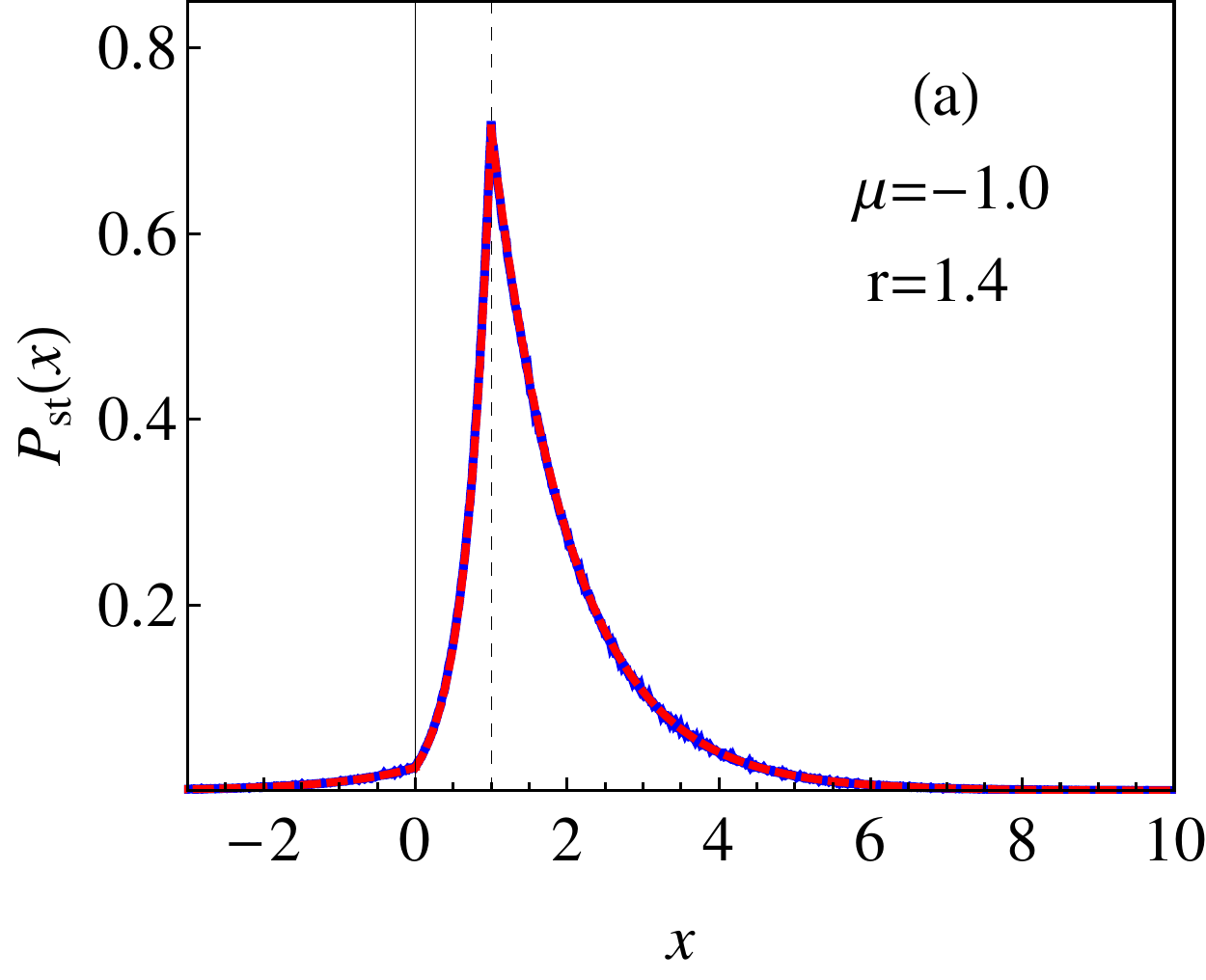}
\includegraphics[width=.3\hsize]{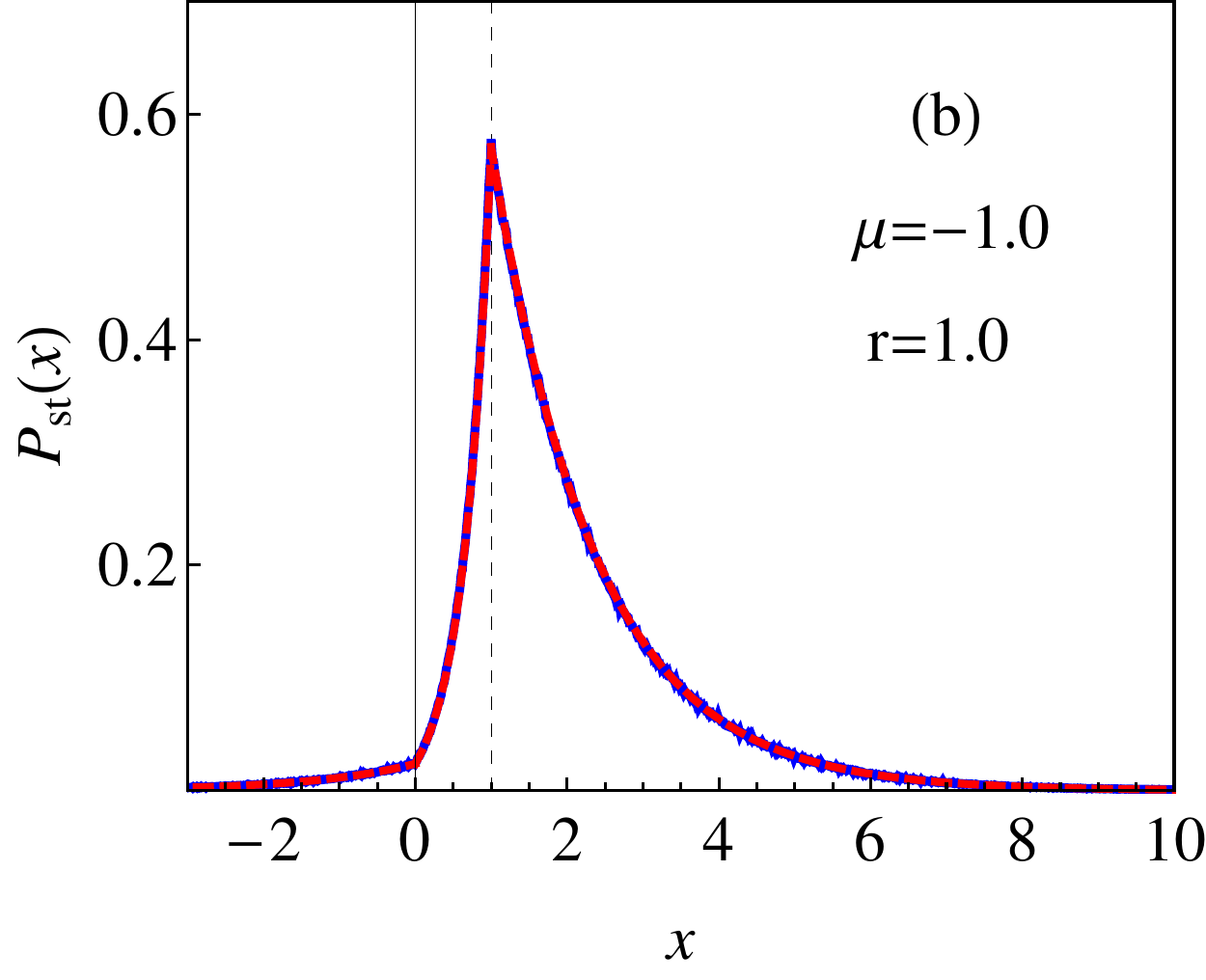}
\includegraphics[width=.3\hsize]{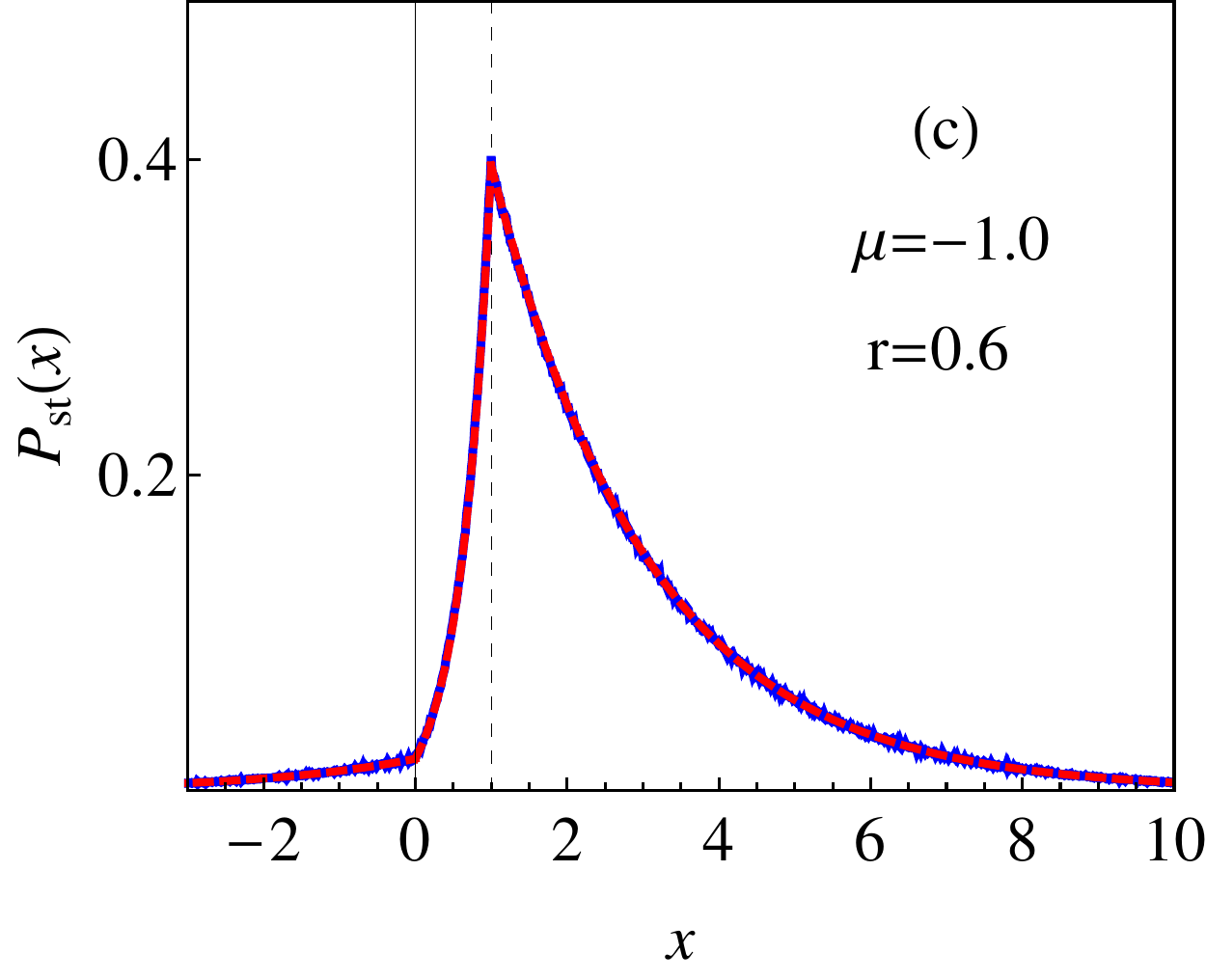}
\caption{(Color online) Stationary distribution $P_{\rm st}(x|x_0)$ for
the unbounded potential $V(x)=\mu|x|$, with $\mu <0$.
We choose $D=0.5,x_0=1.0,\mu=-1.0$ while vary $r$. The (red) dashed
line plots the analytical result for $P_{\rm st}(x)$, while the (blue)
points are numerical simulation results. Also,
the vertical solid and dashed lines indicate the location of the
unstable maximum of the unbounded potential and the reset point
$x_0$ respectively.}
\l{fig2}
\end{figure*}

%%%%%%%%%%%%%%%%%%%%%%%%%%%%%%%%%%%%%%%%%%%%%%%%%%%%%%%
%%%%%%%%%%%%%%%%%%%%%%%%%%%%%%%%%%%%%%%%%%%%%%%%%%%%%%%
\subsection{The case of a quadratic potential}
We now consider the case of a harmonic potential centred
around $0$ which is either its minimum or the maximum.
As before, reset takes place at $x_0$.
One can again identify two regions in $x$, namely, region I
$(x>x_0)$, and region II $(x<x_0)$. We solve
Eq.~(\ref{eq:steady-state}) in each region and use the fact
that the solutions
are continuous at $x=x_0$ while the derivatives are not.
This can be seen by integrating
Eq.~(\ref{eq:steady-state}) over an infinitesimal region
around $x=x_0$ where one finds
\be
\fr{dP_{\rm st}^I(x|x_0)}{dx}\Big|_{x=x_0}-\fr{dP_{\rm
st}^{II}(x|x_0)}{dx}\Big|_{x=x_0}=-\fr{r}{D}.
\l{eq:matching-condition}
\ee
This is consistent with the fact mentioned in \eref{MC1}.
Similar to the last section,
in the following we derive the steady state solutions for
both the stable and the unstable landscape.

%%%%%%%%%%%%%%%%%%%%%%%%%%%
\subsubsection{Bounded potential:~$V(x)=(\mu/2) x^2$}
We first consider the case where $\mu>0$ and this is
the case of a bounded harmonic potential.
The solutions are then given by
\bea
P_{\rm st}^{I}(x|x_{0})&=&c_{1}e^{-\fr{\mu}{2D}x^{2}}
H\Big(-\fr{r}{\mu},\sqrt{\fr{\mu}{2D}}x\Big)\nonumber \\
&&+c_{2}e^{-\fr{\mu}{2D}x^{2}}
{}_{1}F_{1}\Big(\fr{r}{2 \mu};\fr{1}{2};\fr{\mu}{2D}x^{2}\Big), \nonumber \\
P_{\rm st}^{II}(x|x_{0})&=&c_{3}e^{-\fr{\mu}{2D}x^{2}}
H\Big(-\fr{r}{\mu},\sqrt{\fr{\mu}{2D}}x\Big)\nonumber \\
&&+c_{4}e^{-\fr{\mu}{2D}x^{2}}
{}_{1}F_{1}\Big(\fr{r}{2 \mu};\fr{1}{2};\fr{\mu}{2D}x^{2}\Big),
\l{eq:PDF-quadpot}
\eea
where $H(-n,x)$ is the Hermite polynomial of negative order $n$, and
${}_{1}F_{1}(a;b;x)$ is the Kummer confluent hypergeometric function. 
We note that $H(-n,\sqrt{\mu} x)$ converges as
$x^{-n}$ when $x \to \infty$ but diverges as $x^{n-1}e^{\mu x^2}$
when $x \to -\infty$. But ${}_{1}F_{1}(a;b;\mu x^{2})$ is even in $x$ and diverges as
$e^{\mu x^2}x^{a-b}$ when
$x \to \pm \infty$. However, these functions are
multiplied with $e^{-\mu x^2}$ and then the exponentials
cancel each other which makes the additional algebraic
form important at the asymptotic limits. This results
in two distinct situations namely $r \ge \mu$ and $r<\mu$.
In the first case,
one needs to choose $c_2=0$ for the convergence of
the steady state. However, in the second case,
one can show that
it is not necessary to choose $c_2=0$, rather
there are infinite choices for $c_2$ and for each, $c_1$
will be automatically determined by the
normalization condition. In this paper,
we choose $c_2=0$  to
maintain an identical structure between the two cases.

For further analysis, let us choose $D=1/2$, without loss of generality.
It will prove useful to define the following quantities:
\bea
&&z_{1}(r,\mu,x_{0})\equiv\sqrt{\mu}x_{0}~H\Big(-\fr{r}{\mu},\sqrt{\mu}x_{0}\Big)~{}_{1}F_{1}\Big(1+\fr{r}{2
\mu};\fr{3}{2};\mu
x_{0}^{2}\Big)\nonumber \\
&&+H\Big(-1-\fr{r}{\mu},\sqrt{\mu}~x_{0}\Big)~{}_{1}F_{1}\Big(\fr{r}{2
\mu};\fr{1}{2};\mu x_{0}^{2}\Big), \\
&&a_{1}(r,\mu,x_{0})\equiv\sqrt{\mu}e^{\mu
x_{0}^{2}}~{}_{1}F_{1}\Big(\fr{r}{2 \mu};\fr{1}{2};\mu
x_{0}^{2}\Big), \\
&&b_{1}(r,\mu,x_{0})\equiv\sqrt{\mu}e^{\mu
x_{0}^{2}}~H\Big(-\fr{r}{\mu},\sqrt{\mu}~x_{0}\Big).
\eea
Using these definitions and from \eref{eq:matching-condition}, we
get
\bea
&&c_{3}=c_{1}-\fr{a_{1}(r,\mu,x_{0})}{z_{1}(r,\mu,x_{0})},\\
&&c_{4}=\fr{b_{1}(r,\mu,x_{0})}{z_{1}(r,\mu,x_{0})}.
\l{eq:constants}
\eea
Thus, $c_{4}$ is independent of $c_{1}$, while $c_{3}$ depends on
$c_{1}$ and can be evaluated once $c_{1}$ is found from the
normalization condition:
\bea
\int_{-\infty}^{x_{0}}dx~P_{\rm st}^{II}(x|x_{0})+
\int_{x_{0}}^{\infty}dx~P_{\rm st}^{I}(x|x_{0})=1.
\l{eq:Normalization}
\eea
That said, one obtains the full steady state solutions
from \eref{eq:PDF-quadpot}.
%%%%%%%%%%%%%%%%%%%%%%%%%%
\begin{figure*}
\includegraphics[width=.3\hsize]{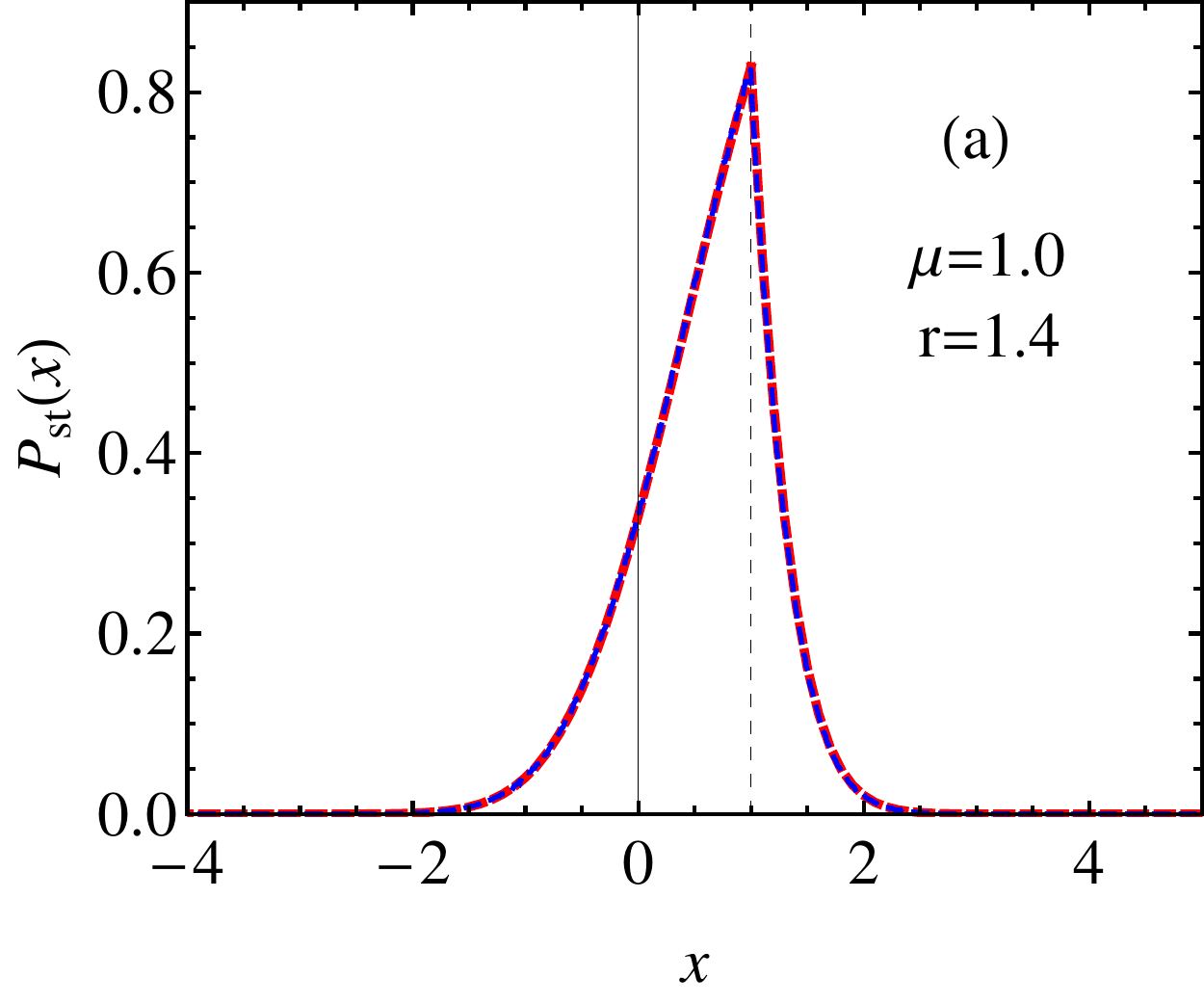}
\includegraphics[width=.3\hsize]{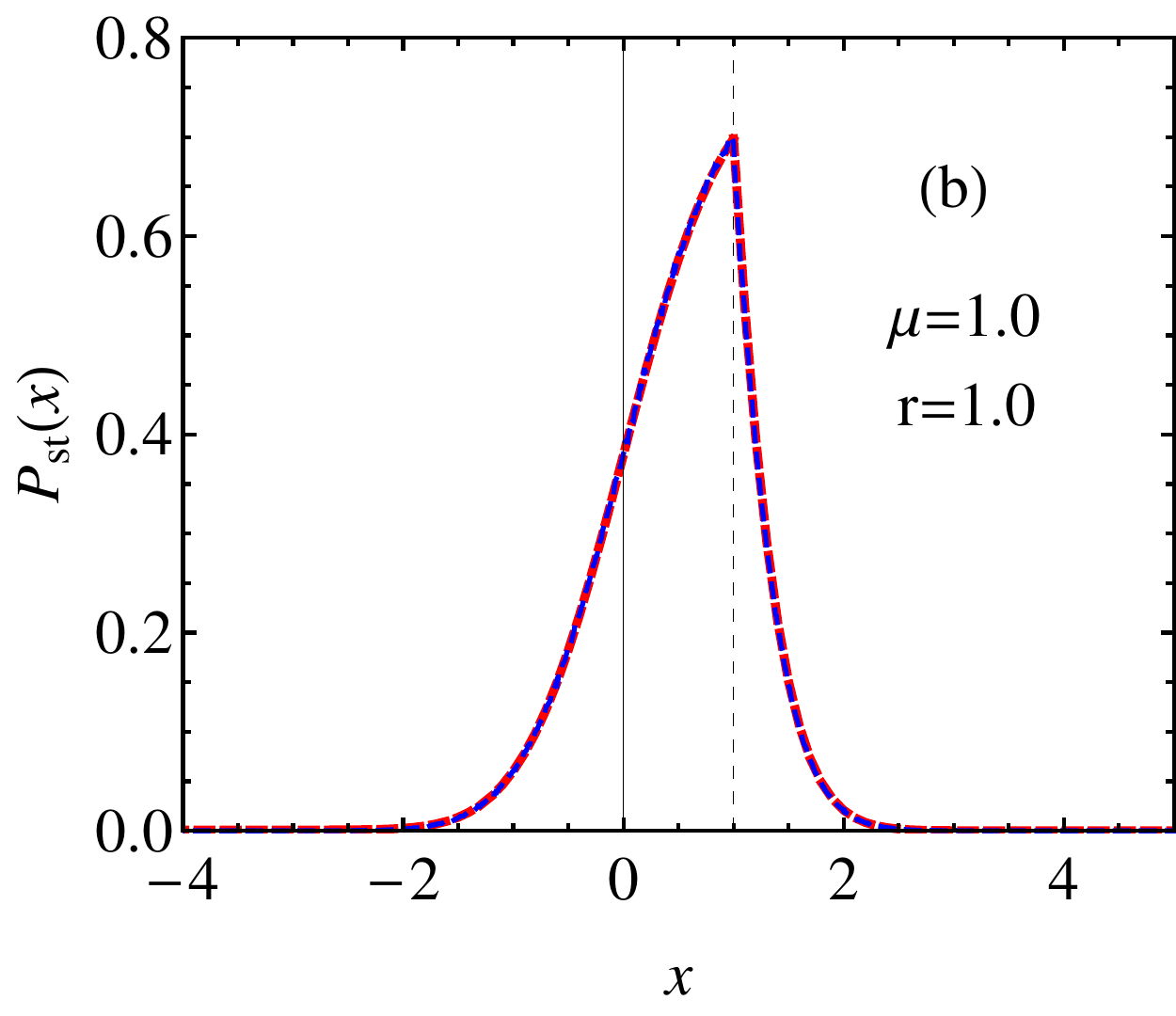}
\includegraphics[width=.3\hsize]{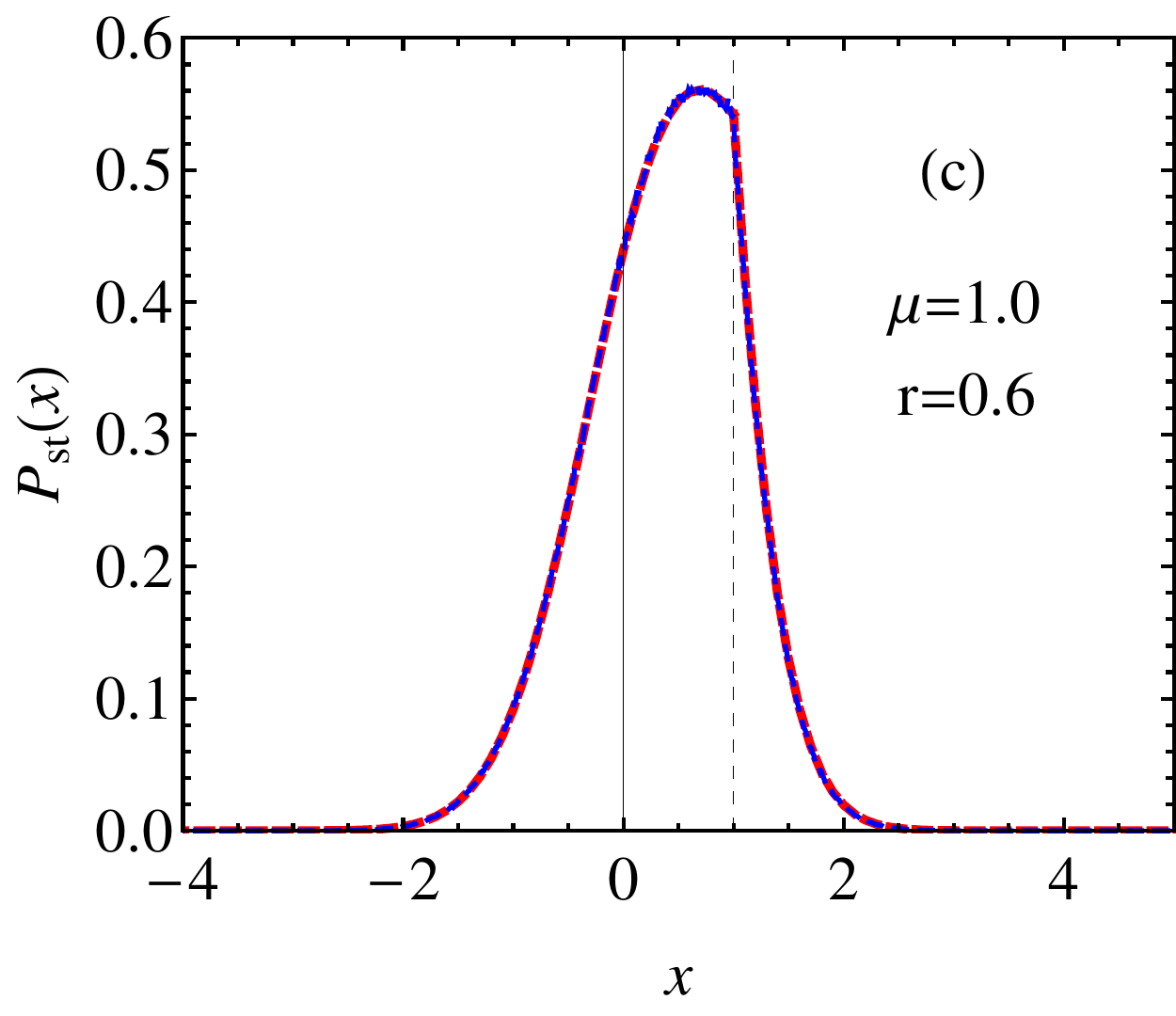}
\caption{(Color online) Stationary distribution $P_{\rm st}(x)$ for
the potential $V(x)=(\mu/2)x^2$, with $\mu >0$.
We choose $D=0.5,~x_0=1.0,~\mu=1.0$ while vary $r$. The (red) dashed
line plots the analytical result for $P_{\rm st}(x)$, while the (blue)
points are numerical simulation results. The
vertical solid and dashed lines indicate the location of the
stable minimum of the bounded potential and the reset point
$x_0$ respectively.
The motion of the peak is also clear from the figure.}
\l{fig3}
\end{figure*}
%%%%%%%%%%%%%%%%%%%%%%%%%%%

\fref{fig3} shows a comparison between simulations and theory for
steady state \eref{eq:PDF-quadpot} demonstrating a very good agreement. 
We note that, there
is only cusp at the reset point $x=x_0$. When $r$ is large compared to
$\mu$, the distribution is peaked around $x=x_0$ with a non-Gaussian
form. However, when $\mu$ is much greater than $r$, we get a
distribution peaked around the minimum of the potential. 
In between, the peak moves between $x_0$ and the minimum.
This generic feature of the distribution is clear from
\fref{fig3}.

%%%%%%%%%%%%%%%%%%%%
\subsubsection{Unbounded potential:~$V(x)=-(\mu/2) x^2$}
We proceed further with a similar analysis in the case of the unbounded harmonic potential
and the solutions are given by
\bea
P_{\rm st}^{I}(x|x_{0})&=&d_{1}
~H\Big(-1-\fr{r}{\mu},\sqrt{\fr{\mu}{2D}}x\Big)\nonumber \\
P_{\rm st}^{II}(x|x_{0})&=&d_{3}
~H\Big(-1-\fr{r}{\mu},\sqrt{\fr{\mu}{2D}}x\Big)\nonumber \\
&&+d_{4}
~{}_{1}F_{1}\Big(\fr{1}{2}+\fr{r}{2 \mu};\fr{1}{2};\fr{\mu}{2D}x^{2}\Big),
\l{eq:PDF-quadpot-unbound}
\eea
where $H(-n,x)$ is the Hermite polynomial of negative order $n$, and
${}_{1}F_{1}(a;b;x)$ is the Kummer confluent hypergeometric function
same as before. Choosing $D=1/2$ and
using the boundary conditions \eref{eq:matching-condition}, one obtains
\bea
&&d_{3}=d_{1}-\fr{a_{2}(r,\mu,x_{0})}{z_{2}(r,\mu,x_{0})},\\
&&d_{4}=\fr{b_{2}(r,\mu,x_{0})}{z_{2}(r,\mu,x_{0})},
\l{eq:constants2}
\eea
where 
\bea
&&z_{2}(r,\mu,x_{0})\equiv(r+\mu)\bigg[\sqrt{\mu}x_{0}~H\Big(-1-\fr{r}{\mu},\sqrt{\mu}x_{0}\Big)~\nonumber \\
&&{}_{1}F_{1}\Big(\fr{3}{2}+\fr{r}{2\mu};\fr{3}{2};\mu x_{0}^{2}\Big)\nonumber \\
&&+H\Big(-2-\fr{r}{\mu},\sqrt{\mu}~x_{0}\Big)~{}_{1}F_{1}\Big(\fr{1}{2}+\fr{r}{2
\mu};\fr{1}{2};\mu x_{0}^{2}\Big) \bigg], \\
&&a_{2}(r,\mu,x_{0})\equiv r \sqrt{\mu}~{}_{1}F_{1}\Big(\fr{1}{2}+\fr{r}{2 \mu};\fr{1}{2};\mu
x_{0}^{2}\Big), \\
&&b_{2}(r,\mu,x_{0})\equiv r \sqrt{\mu}~H\Big(-1-\fr{r}{\mu},\sqrt{\mu}~x_{0}\Big).
\eea
Then $d_{1}$ can be found using the normalization condition \eref{eq:Normalization}
as before and the solutions are deduced from \eref{eq:PDF-quadpot-unbound}.

%%%%%%%%%%%%%%%%%%%%%%
\begin{figure*}
\includegraphics[width=.3\hsize]{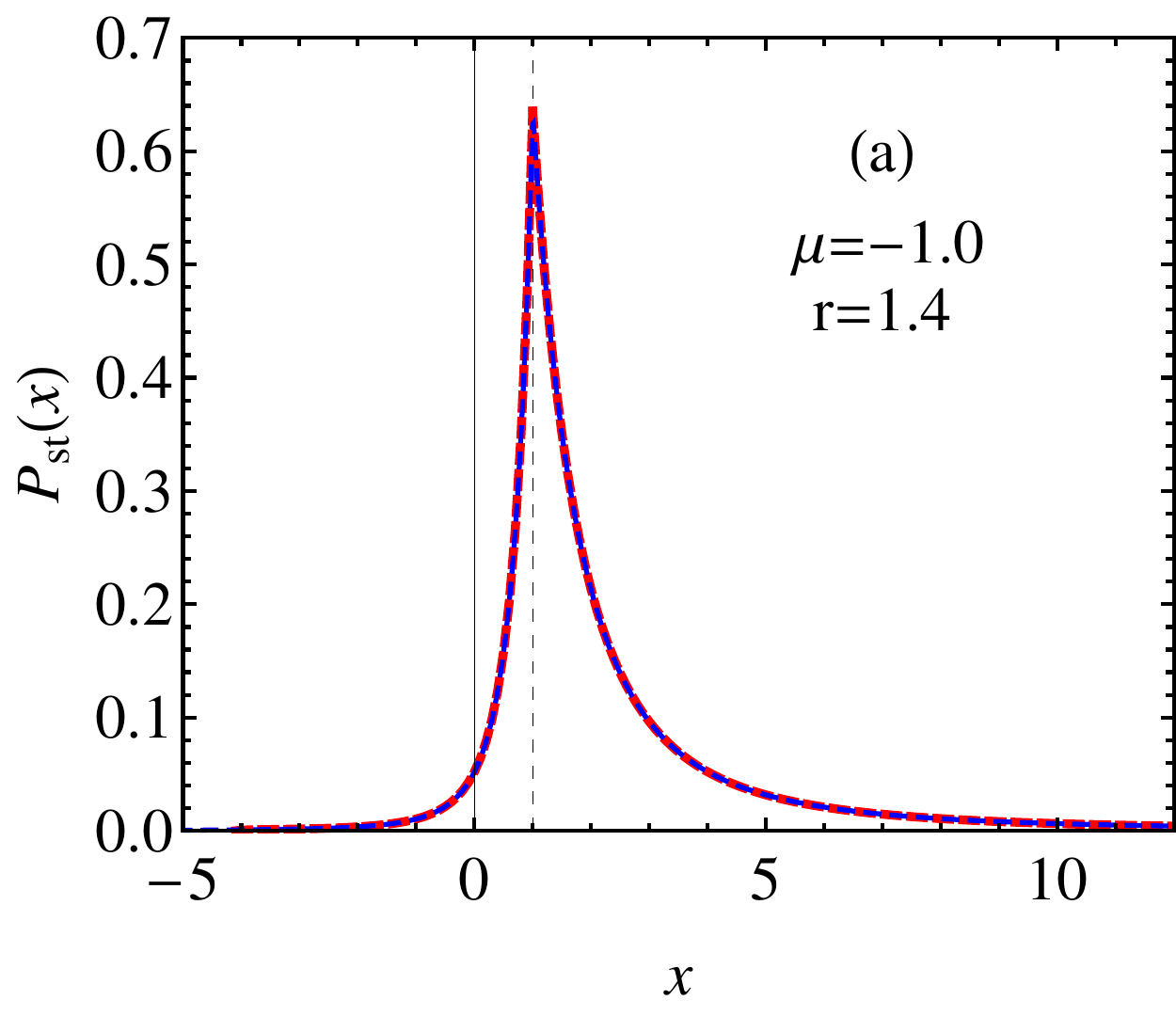}
\includegraphics[width=.3\hsize]{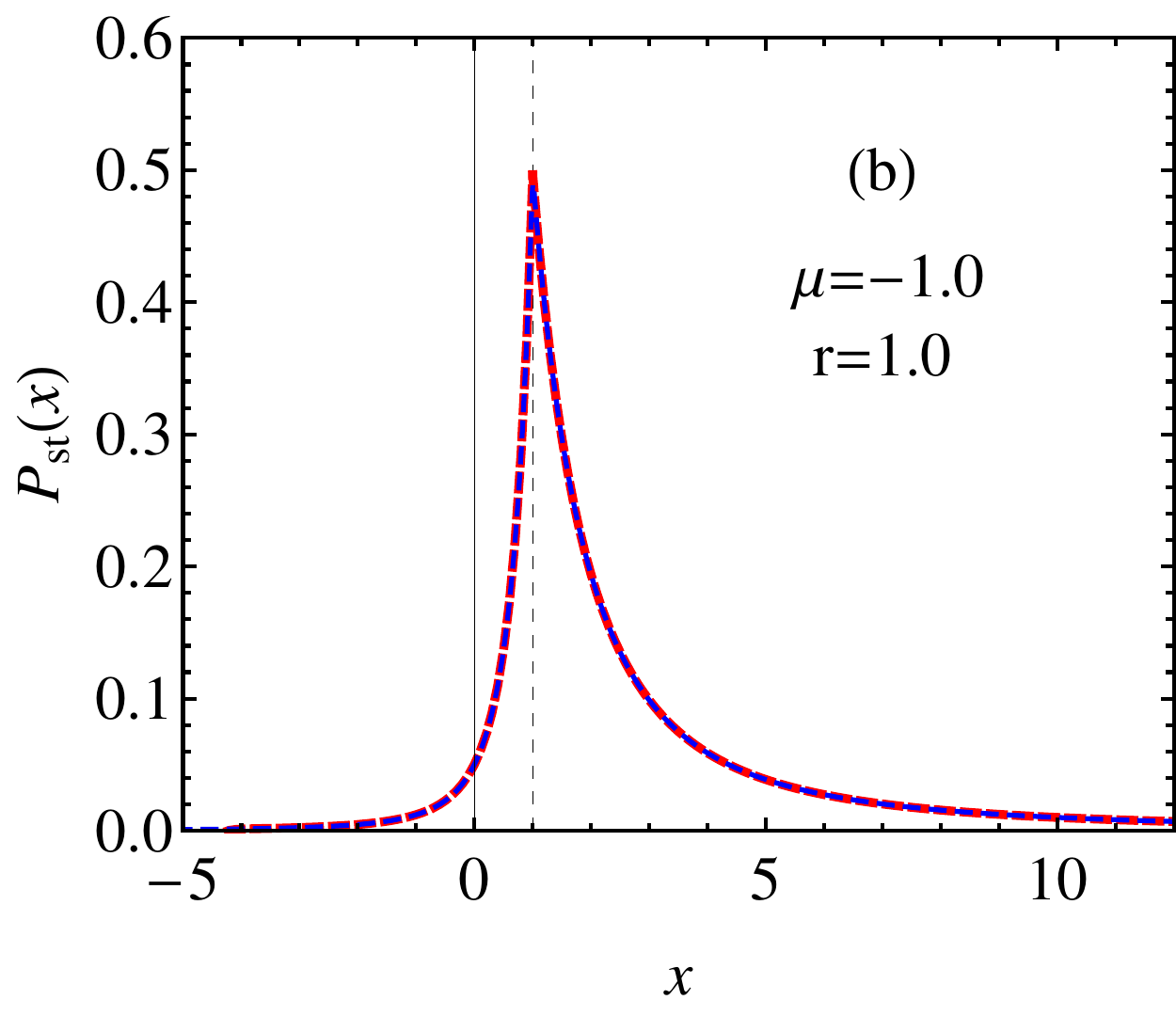}
\includegraphics[width=.3\hsize]{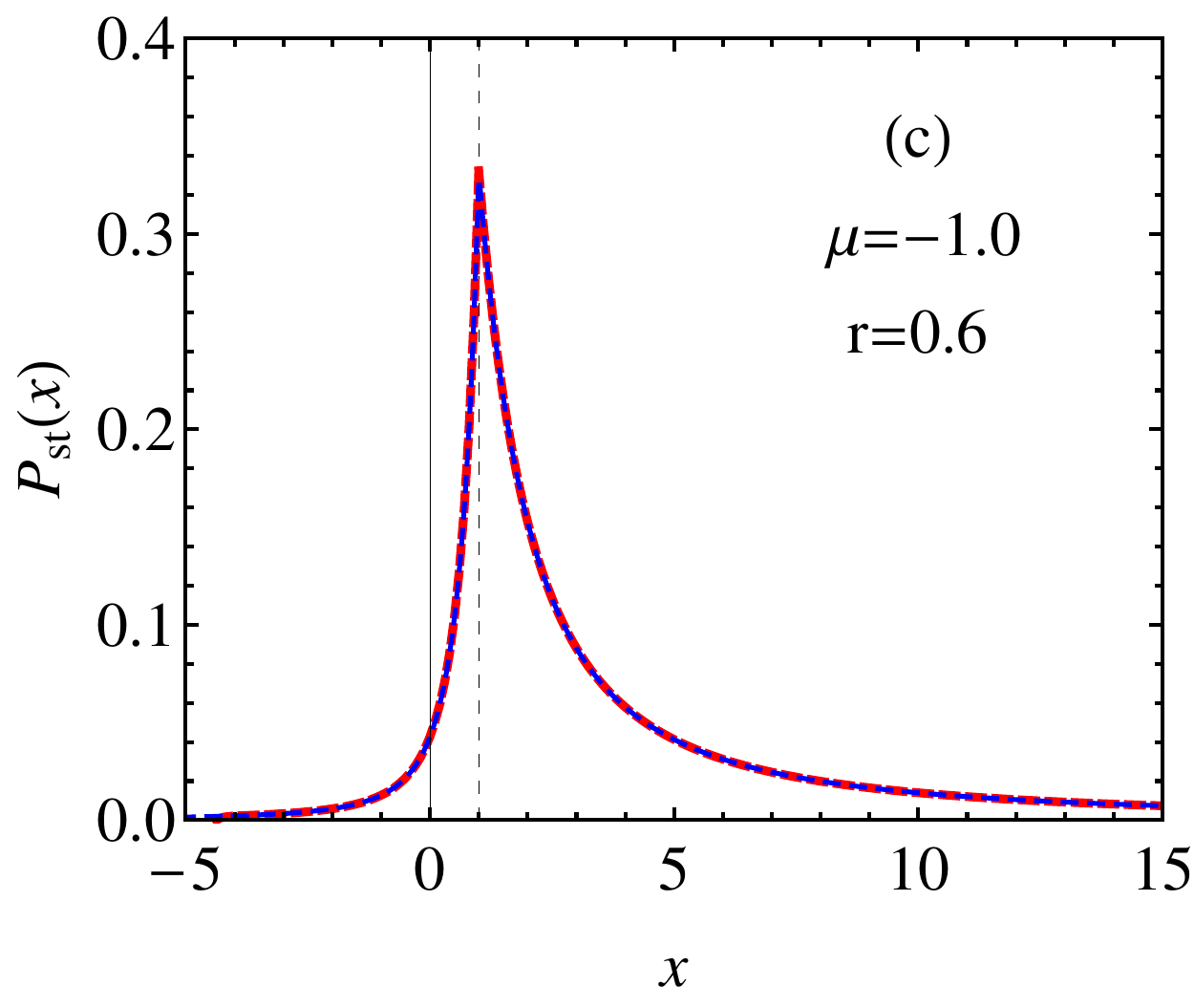}
\caption{(Color online) Stationary distribution $P_{\rm st}(x)$ for
the potential $V(x)=(\mu/2)x^2$, with $\mu <0$.
We choose $D=0.5,~x_0=1.0,~\mu=-1.0$ while vary $r$. The (red) dashed
line plots the analytical result for $P_{\rm st}(x)$, while the (blue)
points are numerical simulation results. The vertical
solid and dashed lines indicate the location of the
unstable maximum of the unbounded potential and the reset point
$x_0$ respectively.}
\l{fig4}
\end{figure*}
%%%%%%%%%%%%%%%%%%%%%%%%

\fref{fig4} shows a comparison between simulations and theory for 
steady state \eref{eq:PDF-quadpot-unbound}, demonstrating a very good agreement. 
We note that, there
is only cusp at the reset point $x=x_0$. When $r$ is large compared to
$\mu$, the distribution is peaked around $x=x_0$ with a non-Gaussian
form. However, when $\mu$ is much greater than $r$, the peak
does not move unlike the case of the bounded potential.
Nevertheless, in this limit, the system takes longer time to reach
the steady state with a peak well set at $x=x_0$
indicating a fat tailed distribution at large $x$ \cite{Arnab-14}.
We refer to the
\fref{fig4} which characterizes this generic feature.

%%%%%%%%%%%%%%%%%%%%%%%%%%%%%%%

\subsection{General $V(x)$: Possible Steady States}
\l{general-Vx}
We generalize our discussion for arbitrary potential that has a form $V(x)= \mu |x|^\delta$. 
When $\mu>0$, that is the potential is stable with minimum
at $x=x_{\text{min}}$, one will always achieve the steady state around $x_{\text{min}}$
irrespective of the resetting. Nevertheless, resetting will invoke the non differentiability 
in the steady state resulting in a cusp at the reset point $x_0\ne x_{\text{min}}$.
Here one can talk about 
two extreme limits: one is
when the strength of the potential is much greater than the reset rate
and one expects 
a steady state solution of form $\sim e^{-V(x)}$ centred around $x_{\text{min}}$
with a small but non vanishing cusp at $x_0$ \eref{MC1}.
In the other limit, when reset rate dominates the potential strength,
one finds a non Gaussian form around $x_0$. However, in between, the 
peak of the steady state moves from $x_{\text{min}}$ to $x_0$ as one varies
 $r$ but keeping $\mu$ fixed. This is a generic feature that can be seen
for any $\delta$.

Now consider the case when $\mu<0$.
There is no stable minimum of the potential,
hence no steady state since the particle escapes to infinity
in the absence of stochastic resetting.
However, we notice that one can have a steady state
when resetting is introduced under certain conditions which we
discuss in the following. We can find a steady
state if and only if $V(x)$ is such that the particle 
starting from $x_0$ does not escape to infinity at a finite time in the 
absence of resetting. Note that the escape time is given by
 $t_{\text{esc}}=-\int_{x_{0}}^{\infty}~[V'(x)]^{-1}~dx=[x_0^{\delta-2} (\delta-2)\delta \mu]^{-1}$ 
for $\delta>2$. On the other hand, the waiting 
time distribution for resetting is given by Poisson distribution namely $re^{-r \tau}$,
with the average time between two resets is simply given by $t_{\text{reset}}=1/r$,
which is always finite. It is then obvious that if $t_{\text{esc}}<t_{\text{reset}}$
the particle always escapes and there is no steady state. However,
one indeed achieves a steady state if $t_{\text{esc}}>t_{\text{reset}}$ 
even for $\delta>2$. This can be realized by increasing the reset
rate so that it gets reset promptly before escaping. On the contrary,
for $\delta \le 2$, one finds $t_{\text{esc}} \to \infty$, thus 
always maintaining
a steady state through resetting.
We have discussed the cases of $\delta=2$ (harmonic) and $\delta=1$
(mod) for both positive and negative $\mu$ in
great details. For positive $\mu$, the steady states
and the motion of the
peak as well is followed from the \fref{fig1}(b)-(c),
\fref{fig3}(b)-(c). But for negative $\mu$,
the peak is always set at $x_0$ indicating the fact that
the steady state is solely due to the reset mechanism.
This is realized from \fref{fig2}, \fref{fig4}.

%%%%%%%%%%%%%%%%%%%%%%%%%%%%%%%%%%%%%%%%%%%%%%%%%%%%%%%
\section{relaxation to the steady state}
\l{transient}
In this section, we investigate the transient behavior of
the stochastic resetting mechanism. We recall that
the particle starts at $x=x_0$ at $t=0$ and finally attains
a steady state either at $x=x_0$ or $x=x_{\text{min}}$ as $t\to\infty$
depending on the potential landscape.
In between, the particle
position PDF shows rich behavior which can be quantified
by studying the relaxation dynamics of the propagator.
We first recall \eref{transient} 
\be
\frac{\partial P}{\partial
t}=D\frac{\partial^{2}P}{\partial
x^{2}}+\frac{\partial [V'(x)P]}{\partial
x}-rP+r\delta(x-x_0),
\ee
with the boundary conditions $P(x\to \pm \infty,t)=0$ and
the initial condition $P(x,t=0)=\delta(x-x_0)$.
Now to characterize the transient states, one has to
solve \eref{transient} for the time dependent
propagator. To do this, we first separate
$P(x,t)=f(x)+b(t,x)$ where, $f(x)$ gives the steady state
solution and $b(t,x)$ describes the relaxation towards it.
As a representative case, we choose the free diffusion with
no potential.
The steady state solution $f(x)$ then satisfies the simple
equation \eref{eq:steady-state} with the boundary conditions
$f(x\to \pm \infty)=0$,
\bea
Df''(x)-rf(x)+r\delta(x-x_0)=0,
\eea
and this gives rise to the solution 
\bea
f(x)=\fr{\alpha}{2}\exp\big[-\alpha|x-x_0|\big],
\l{steady-state-1}
\eea
where $\alpha=\sqrt{\fr{r}{D}}$  is an inverse length
scale denoting to the typical distance
diffused by the particle between the
resets \cite{Evans:2011-1}.
The time dependent part is given by
\bea
\partial_{t}b(t,x)=D\partial_{x}^{2}b(t,x)-rb(t,x),
\l{transient-1}
\eea
\\
with the boundary conditions $b(t,x\to \pm \infty)=0$ and
$b(t\to\infty,x)\to 0$. The initial condition
is given by $b_0(x)\equiv b(t=0,x)=P(x,0)-f(x)$.
This results in the complete form of the relaxation term
given by
\bea
b(t,x)&=&e^{-rt}~\fr{\exp\big[-\fr{(x-x_0)^2}{4Dt}\big]}{\sqrt{4\pi Dt}}-\fr{\alpha}{2}\cosh\big[-\alpha(x_0-x)\big]\nonumber \\
&+&\fr{\alpha}{4}~\exp\big[-\alpha(x_0-x)\big]~\text{erf}\bigg[\fr{x-x_0+2Dt\alpha}{\sqrt{4\pi Dt}}\bigg]\nonumber \\
&+&\fr{\alpha}{4}~\exp\big[\alpha(x_0-x)\big]~\text{erf}\bigg[\fr{-x+x_0+2Dt\alpha}{\sqrt{4\pi Dt}}\bigg].
\l{transient-2}
\eea
\\
Now, \eref{steady-state-1} and \eref{transient-2} constitute the 
full propagator.
We refer to the top panel of \fref{relaxation-1} which specifies the relaxation for
this particular case. A similar analysis can also be made for a Brownian
particle diffusing in a potential in the presence of resetting.
For instance, we analyze the case of a bounded harmonic potential
$V(x)=(\mu/2)x^2$
with minimum $x_\text{min}=0$ while the reset point is at $x_0\ne 0$.
This gives rise to a competition between the potential and the
reset mechanism thus reaching a steady state
as discussed in \sref{general-Vx}. The bottom panel
of \fref{relaxation-1} characterizes the transient
behavior with respect to $t$ for $\mu=1.0,~r=0.6$.
We also notice that the steady state achieved at the end
is identical with that obtained in \fref{fig3}(c).

%%%%%%%%%%%%%%%%%
\begin{figure}
\includegraphics[width=8cm]{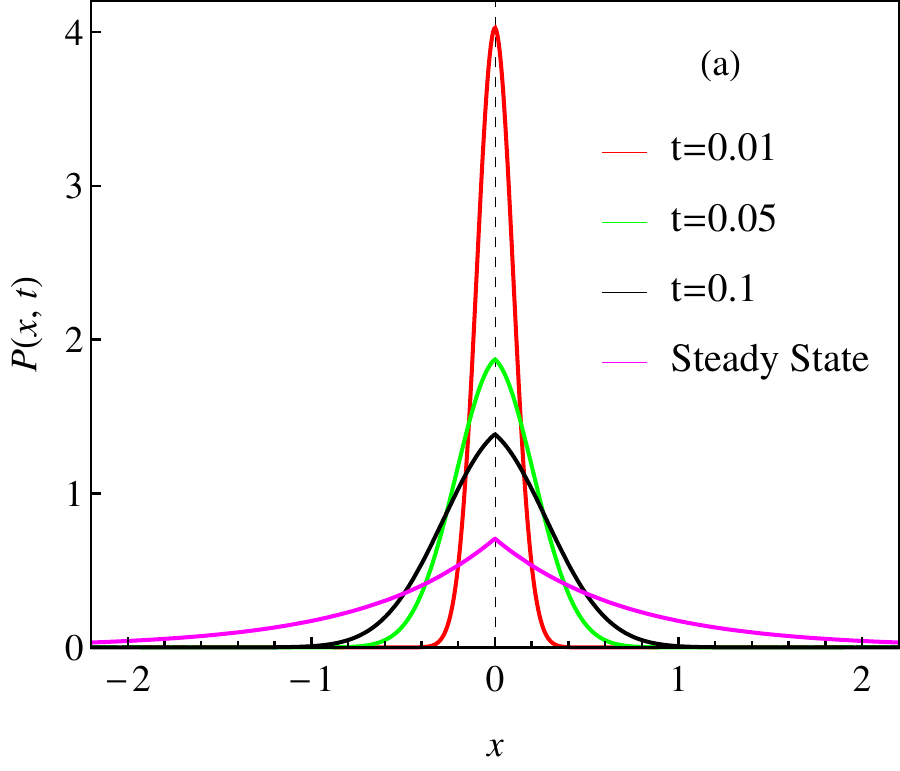}
\includegraphics[width=8cm]{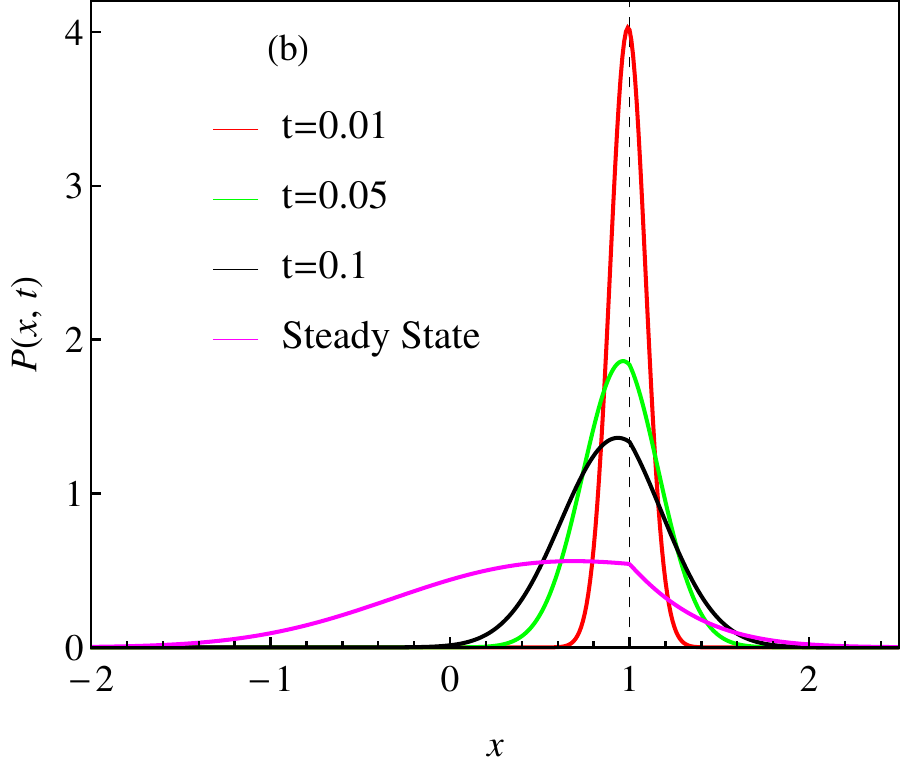}
\caption{(Color online) The time dependent propagator for the
free and forced diffusion case in the presence of stochastic resetting
has been plotted. It depicts the transient of the propagator
from time zero to its steady state. The top panel (a) and
the bottom panel (b) describe
the free and the forced diffusion case respectively.
The parameters are chosen to be $r=0.6,~\mu=1.0$
and $D=0.5$.
The vertical dashed line
marks the reset position $x_0$.
In the free case (a) we consider $x_0=0$,
but for the case (b) we set $x_0=1.0$, which
is different from the stable minimum $x_\text{min}=0$.}
\l{relaxation-1}
\end{figure}
%%%%%%%%%%%%%%%%%%

\section{Summary}
\l{summary}
In this work, we have considered
a Brownian particle diffusing in
an arbitrary potential landscape in the presence
of the stochastic resetting mechanism. We have investigated
the steady state properties of the position distribution
of the particle for two representative choices of the
potential namely the mod and the harmonic potential.
It has been shown that the steady states
have distinct differences depending on the nature of the potential.
We also derive the conditions for the existence of the
steady state for any potential landscape of higher order.
Also we have realized the transient behavior of
the propagator approaching to the steady state.
We have studied two representative cases
in this context though the extension to higher
order potential does not offer more physical insights.

Furthermore, resetting has been found
to have a profound consequence
on the first passage properties of a diffusing particle.
In recent times, there have been extensive studies
on this to have a
discreet idea
not only restricted to one dimension but to higher dimension
as well \cite{Evans:2014}. Consequently, the study of two observables
namely the local time and the occupation time
turns out to be very useful to understand the mechanism near the
reset point. Local time measures the time that the process
visits a reference point (which is basically the reset point)
while the residence time or the
occupation time measures the time that the process
stays above that point \cite{Majumdar:2005, Sabhapandit:2006}.
These observables show rich
behavior when the resetting dynamics is combined \cite{Arnab-14}.
There are lot of open questions in the context of
stochastic resetting mechanism. One can generalize
resetting to the systems where the resetting
takes place to a region instead of a reference
point at a constant rate. Also exploring the span or the
extremum (namely maximum or minimum) of a dynamics
under the resetting paradigm will be very interesting in the
connection with the extreme value statistics \cite{Satya-14}.
\\
%%%%%%%%%%%%%%%%%%%%%%%%%%%%%%%%%%%%%%%%%%%%%%%%%%%%%%%
\section{Acknowledgements}
The author thanks
Satya N. Majumdar, Sanjib Sabhapandit and Shamik Gupta
for useful
discussions.
The author also thanks the Galileo Galilei Institute for Theoretical
Physics, Florence, Italy for the hospitality and the INFN for partial
support during the completion of this work.
%%%%%%%%%%%%%%%%%%%%%%%%%%%%%%%%%%%%%%%%%%%%%%%%%%%%%%%

%%%%%%%%%%%%%%%%%%%%%%%%%%%%%%%%%%%%%%%%%%%%%%%%%%%%%%%

\begin{thebibliography}{1}


\bibitem{Bell-91} W. J. Bell, Searching behaviour: the behavioural ecology of finding resources, (Chapman and Hall,
London 1991).

\bibitem{Benichou-11} O. Benichou, C. Loverdo, M. Moreau, and R. Voituriez, Rev. Mod. Phys. \textbf{83}, 81 (2011).

\bibitem{Adam-68} G. Adam and M. Delbruck, Reduction of dimensionality in biological diffusion processes, in
Structural Chemistry and Molecular Biology, A. Rich and N. Davidson Eds. (W.H. Freeman
and Company, San Francisco; London, 1968).

\bibitem{Montanari-02} A. Montanari and R. Zecchina, Phys. Rev. Lett. \textbf{88}, 178701 (2002).


\bibitem{Bartumeus-09} F. Bartumeus and J. Catalan, J. Phys. A:Math. Theor. \textbf{42}, 434002 (2009).

\bibitem{Cates-12} M.E.Cates, Rep. Prog. Phys. \textbf{75}, 042601 (2012).

\bibitem{Manrubia-99} S. C. Manrubia and D. H. Zanette,  Phys.Rev. E \textbf{59}, 4945 (1999).

\bibitem{Visco-10} P. Visco, R. J. Allen, S. N. Majumdar, and M. R. Evans,  Biophysical Journal \textbf{98}, 10991108 (2010).

\bibitem{Montero-13} M. Montero and J. Villarroel,  Phys. Rev. E \textbf{87}, 012116 (2013).


\bibitem{Evans:2011-1} M. R. Evans and S. N. Majumdar, Phys. Rev. Lett. \textbf{106},
160601 (2011).

\bibitem{Evans:2011-2} M. R. Evans and S. N. Majumdar, J. Phys. A: Math.
Theor. \textbf{44}, 435001 (2011).

\bibitem{Evans:2013} M. R. Evans, S. N. Majumdar, and K. Mallick, J. Phys.
A: Math. Theor. \textbf{46}, 185001 (2013).

\bibitem{Whitehouse:2013} J. Whitehouse, M. R. Evans, and S. N. Majumdar, Phys.
Rev. E \textbf{87}, 022118 (2013).

\bibitem{Henkel:2014} X. Durang, M. Henkel, and H. Park, J. Phys.
A: Math. Theor. \textbf{47}, 045002 (2014).

\bibitem{Gupta:2014} S. Gupta, S. N. Majumdar, and G. Schehr, Phys. Rev.
Lett. {\bf 112}, 220601 (2014).

\bibitem{Evans:2014} M. R. Evans and S. N. Majumdar, arXiv:1404.4574 (2014).


\bibitem{Majumdar:2005} S. N. Majumdar, Curr. Sci. {\bf 89}, 2076 (2005).

\bibitem{Sabhapandit:2006} S. Sabhapandit, S. N. Majumdar, and A. Comtet, Phys.
Rev. E\textbf{ 73}, 051102 (2006).

\bibitem{Arnab-14} A. Pal, P. Basu, and S. Gupta, in preparation.

\bibitem{Satya-14}
S. N. Majumdar and A. Pal, arXiv:1406.6768 (2014).


\end{thebibliography}
\end{document}